\documentclass[preprint]{aastex}

\shorttitle{Stern et al.}
\shortauthors{Argon in Comet Hale-Bopp}

\begin{document}


\title{The Discovery\\ of Argon in Comet C/1995 O1 (Hale-Bopp)}


\author{S.A. Stern} 
\affil{Space Studies Dept., Southwest Research Institute, Boulder, CO 80302}
\email{astern@swri.edu}

\author{D.C. Slater}
\affil{Space Science Dept., Southwest Research Institute, San Antonio TX 78256}

\author{M.C.~Festou\altaffilmark{1}} 
\affil{Space Studies Dept., Southwest Research Institute, Boulder, CO 80302}

\author{J.Wm.~Parker}
\affil{Space Studies Dept., Southwest Research Institute, Boulder, CO 80302}

\author{G.R.~Gladstone}
\affil{Space Science Dept., Southwest Research Institute, San Antonio TX 78256}

\author{M.F.~A'Hearn}
\affil{Dept. of Astronomy, University of Maryland, College Park, MD 20742}

\and

\author{E.~Wilkinson}
\affil{CASA, University of Colorado, Boulder, CO 80302}


\altaffiltext{1}{present address: Observatoire Midi-Pyren\'ees, Toulouse, France.}


\begin{abstract} 

On 30.14 March 1997 we observed the EUV spectrum of the bright comet C/1995 O1
(Hale-Bopp) at the time of its perihelion, using our EUVS sounding rocket
telescope/spectrometer.  The spectra reveal the presence H Ly $\beta$, O$^+$, and, most
notably, Argon.  Modelling of the retrieved Ar production rates indicates that comet
Hale-Bopp is enriched in Ar relative to cosmogonic expectations.  This in turn indicates
that Hale-Bopp's deep interior has never been exposed to the 35--40 K temperatures
necessary to deplete the comet's primordial argon supply.

\end{abstract}




\section{Introduction}

As well preserved, ancient relics of the chemistry present at the formation stage of the
outer solar system, comets provide one of the most valuable tools available for
understanding the formation processes and conditions extant when the planets formed,
$\approx$4.5 Gyr ago \citep{Mea93}.

Among the most long-standing observational goals for understanding cometary origins has
been the search for cometary noble gases.  Owing to their combination of high volatility and
disaffinity to chemical reactions, noble gases provide a key diagnostic to the thermal
history of cometary ices.  More specifically, the members of the He, Ne, Ar, Kr sequence
display successively higher sublimation temperatures; as such, they provide a series of
thermometers that can be exploited to constrain the thermal history, and therefore the
sites of cometary origins.  Although the interpretation of noble gas abundances in
cometary comae is complicated by the details of their trapping and release efficiencies
in cometary ice \citep{Oea91}, their detection has nonetheless been highly desired.
Unfortunately, however, although significant upper limits revealing He 
depletions of $\sim$10$^4$ in comet Austin \citep{Sea92} and Ne depletions of
$\sim$25 in comet Hale-Bopp \citep{Kea97} have been obtained, no detection of
any cometary noble gas emmanating from within a comet has previously been obtained.

He, Ne, Ar, and Kr each have resonance transitions in the far and extreme ultraviolet.
Among the noble gases, argon offers a particularly good combination of comparatively
high cosmogonic abundance, moderate sublimation temperature, and good UV resonance
fluorescence efficiency; together these properties suggested some time ago \citep{Sea92,
Mea93} that argon may be the easiest noble gas to detect in comets.  We took
advantage of the apparition of the unusually active and bright comet C/1995 O1 (Hale-Bopp)
in order to conduct a new and more sensitive search for argon in comets.

\section{The 950--1100 \AA\ Spectrum of Comet Hale-Bopp}

With the objectives of (i) obtaining a general survey spectrum of Hale-Bopp in the EUV
near its perihelion, and (ii) making a more sensitive search for argon than had
previously been attempted for any comet, we launched the EUVS sounding rocket
telescope/spectrometer payload on a suborbital mission timed to coincide with the
comet's perihelion.  This timing also corresponded closely to the epoch of
Hale-Bopp's peak in activity.

The 184 kg EUVS payload \citep{Sea95} consists of a 40-cm diameter grazing incidence
telescope, a long slit Rowland circle spectrograph, and its accompanying Ranicon 2-D
microchannel plate detector, power system, and telemetry electronics.  The telescope is
a diamond-turned f/15 Wolter type II grazing incidence design, with a 30-cm aperture
\citep{Cea89}.  The primary mirror is Ni coated; the secondary is SiC coated.  For the
Hale-Bopp flight, EUVS was configured to study the bandpass from 820--1100 \AA; the
characteristic effective area of the instrument in this bandpass is 0.5 cm$^2$.

EUVS was launched to observe Hale-Bopp on a NASA Black Brant IX sounding rocket from
White Sands, New Mexico, at 03:25 UT on 30 Mar 1997.  At this time Hale-Bopp was less
than 48 hours from perihelion, with a geocentric distance of $\Delta$=1.34 AU, a
heliocentric distance of $R$=0.915 AU, and a heliocentric radial velocity of
$\dot{R}$=--1.20 km s$^{-1}$.  The launch occurred in darkness, with the Sun and the
comet 116 deg and 76.1 deg from the zenith, respectively.  The launch vehicle performed
nominally, lifting the instrument to a peak altitude of 314 km.  During the flight
EUVS remained in the Earth's shadow at all times; the line of sight to the comet
remained above local horizontal.  These two factors effectively eliminated telluric
dayglow emissions from the spectra that EUVS obtained.  The data we describe below were
obtained during the 195 sec period when the payload was above 200 km, where telluric
absorption of the EUV is also negligible \citep{PA92}.

EUVS successfully obtained spectra to survey Hale-Bopp in the EUV bandpass using both its
130$\times$38.6 arcsec (12.5 \AA) medium resolution EUVS (MRES) and 130x6.4 arcsec high
resolution (2 \AA) EUVS (HRES) slits; the centers of these two slits are located 130
arcsec apart.  Based on count rate measurements received in real time as telemetry, we
elected to maximize SNR at the expense of spectral resolution and place the center of
brightness of Hale-Bopp's coma in the MRES; the adjacent HRES slit was therefore located
some 1.27$\times$10$^5$ km away, along the tail axis of the comet.  The EUVS Hale-Bopp
dataset contains both strong H Ly $\beta$, and the first ever spectroscopic detections of
other neutral H and O features in the EUV, cometary O$^+$, and cometary argon
\citep{Sea98}.  In this Letter, we concentrate on the 950--1100 \AA\ portion of the EUVS
data (see Figure 1) in order to focus on the evidence for, and implications of, Ar in
Hale-Bopp; a future publication will discuss the other emissions in the EUVS Hale-Bopp dataset. 

The EUVS wavelength scale was established by a least-squares fit to a series of Pt-lamp
lines imaged onto the detector before and after the flight; these pre- and post-flight
wavelength calibrations concurred to within 3 \AA.  During data reduction, accurate
co-registration of the wavelength scales for the MRES and HRES data was achieved using
the centroid of the prominent 1025.7 \AA\ Lyman $\beta$ emission line in Hale-Bopp.  The
effective area of EUVS was calibrated in flight as a function of wavelength using the
UV-bright B0.5 IV star, $\epsilon$ Per.  A post-flight laboratory effective area
calibration was also obtained, using an O$^+$/Ar resonance source at five wavelengths
across the EUVS bandpass.  The two calibrations are in good ($\pm$35\%) agreement.

Using the effective area of the instrument and the solid angle of the slit(s), we converted
the Hale-Bopp raw counts spectra to the brightness spectra shown in Figure 1.  The brightness error bars
shown in Figure 1 were computed at each plotted wavelength from the standard deviation of
the mean of the uncertainties due to counting statistics (including the term due to the
nominal background subtraction), but neglect the systematic uncertainty in effective area;
errors due to ephemeris and pointing geometry uncertainties ($<<$1 deg) are completely
negligible.

The MRES data in Figure 1 reveals what we believe are the 1048 and 1066 \AA\ ArI features
blended together just redward of the far stronger, 1026 \AA\ HI/OI blend.  Additional
emissions are seen between 975--990 \AA\ and 1080--1097 \AA.  Although these emissions are
clearly statistically significant, their identification is not presently secure; therefore,
the analysis of these emissions and the marginally optically thick H Ly $\beta$ signal EUVS
detected will be discussed in a second publication.  Regarding the latter, here we
simply note that the Lyman $\beta$ feature is blended with a contribution from an OI
[3d$^3$D$_{1,2,0}$--$>$2s$^2$2p$^4$ $^3$P$_{2,1,0}$] triplet at 1025.7, 1027.4, and 1028.7
\AA, which, owing to radiative pumping by Ly $\beta$, is difficult to precisely model, but
which to first order contributes $\sim$15 R; the telluric background Ly$\beta$ brightness
is estimated to contribute some 20 R above the background.  Sky background spectra obtained
while EUVS was maneuvering toward Hale-Bopp contain only a weak instrumental background and
H I Ly $\beta$ emission generated by a combination of geocoronal and interplanetary medium
hydrogen. The fact that no other features are present in EUVS sky background 
spectra, combined with the observing altitude and geometry discussed above, 
virtually eliminates the possibility of the features seen at Hale-Bopp as being due to
telluric contamination.

Returning now to the main subject of this paper, additional evidence for the Ar
identification comes in three forms.  First, the width of the MRES feature extending from
1045--1070 \AA\ is consistent with a pair of blended lines 19 \AA\ apart (such as the two
Ar lines).  Second, the sense of the asymmetry seen in the 1045--1070 \AA\ emission is
consistent with the fact that the 1048 \AA\ line's resonance fluorescence efficiency in
sunlight (i.e., its g-factor) is 2.4 times higher than the 1066 \AA\ line's.  Third, as
shown in the righthand panel of Figure 1, there is also both clear evidence for a 1048
\AA\ Ar feature (present some 3.2 $\sigma$ above the local background) and an hint of the
1066 \AA\ feature in the higher-resolution HRES data, thereby all but eliminating 
pathological detection cases (e.g., due to flat field effects or count statistics
variations) of the Ar features in the main slit.

Since the MRES data has considerably better count levels than the HRES data, we used our
line detection software to retrieve individual, background-subtracted gaussians (with
widths established from calibration line sources) for the two suspected MRES dataset
argon features.  This software, which obtains best fit wavelengths and integrated
brightnesses gaussians for features detected above a specified background, retrieved
brightnesses and statistical error bars of 24$\pm$8 R centered at 1049 \AA, and 12$\pm5$
R at 1066 \AA\ (1 R=10$^6$ ph cm$^{-2}$ s$^{-1}$ emitted into 4$\pi$ sr).  It is worth
pointing out that the 2.0$\pm$0.6 ratio of these two retrieved brightnesses corresponds
well (within our statistics) to the 2.4:1 value predicted by their g-factors, and line
formation theory for optically thin emission.\footnote {The adopted background level we
are using is based on the entire 825--1100 \AA\ MRES spectrum.  If we instead adopt an
upward sloping background that better corresponds to a slope set by the three lowest
pixel brightnesses in the 950--1100 \AA\ MRES spectrum, then one retrieves 22 R at 1049
\AA, and 10 R at 1067 \AA.  Although we do not advocate such a choice of background, we
do note that it does more closely reflect the 2.4:1 brightness ratio expected on the
basis of the g-factors of the two Ar I features.}

\section{Argon and Its Implications}

In order to interpret the Ar emissions, we first converted the signal brightness $B$
in Rayleighs into a slit-average column density $\bar{N}$ along the line of sight
for each Ar line.\footnote {On the date of our observations, the projected MRES slit
size at the distance of the comet was $\approx$37,600$\times$126,600 km.  The Ar
photoionization scale length at 0.915 AU ($\approx$2.7$\times$10$^{6}$ km) is large
compared to the EUVS slit.}  The conversion to column densities in the slit was
performed according to the optically thin approximation:  $\bar{N_i} = (10^6
B_i)/g_i$, where $B_i$ is the brightness of line $i$, and $g_i$ is the photon
fluorescence rate (i.e., ph s$^{-1}$) of that emission at the comet.  We adopted
resonance fluorescence efficiencies at 1 AU of 5.3$\times$10$^{-8}$ s$^{-1}$ and
2.2$\times$10$^{-8}$ s$^{-1}$ for the 1048 and 1066 \AA\ lines (after adjustment for
the variation in solar FUV using the daily F10.7 solar flux index at the time of our
flight)\footnote {Owing to the fact that we observed Hale-Bopp essentially at
perihelion, when its heliocentric radial velocity was consequently small, no Swings
effect correction was made.}; these g-values (derived from \citep{M91, Pea98} were
then adjusted to reflect Hale-Bopp's 0.915 AU heliocentric distance.

To derive an Ar production rate ($Q_{Ar}$), we constructed a simple model of the coma's Ar
distribution, assuming a spherically symmetric, steady-state radial outflow diverging from a
point source.  To calculate Q$_{Ar}$, we assumed an argon outflow velocity in equilibrium
with the H$_2$O outflow, i.e., $v_i$=1.25 km s$^{-1}$ at 0.915 AU.  Based on the MRES slit
brightnesses (and counting+background brightness errors) of the two argon
lines, an error-weighted average Ar production rate of
1.1$\pm$0.3$\times$10$^{29}$ s$^{-1}$, was derived.

To interpret the physical meaning of the EUVS Ar production rate obtained above, we ratio
the derived Ar production rates to the O production rate of the comet.  H$_2$O is the
dominant O-bearing molecule in cometary ices, so we derived the O production rate $Q_O$ on
the established 1.1$\times$10$^{31}$ s$^{-1}$ perihelion H$_2$O production rate
\citep{Bea97, Cea97}.  To account for the oxygen in other O-bearing species (CO, CO$_2$,
CHONs, SiO$_x$) when deriving the Ar production ratio [Ar/O], we adopt $Q_{\rm
oxygen}=1.5Q_{\rm H_2O}$ (based on Giotto NMS data adjusted for the higher CO abundance of
Hale-Bopp; W.~Huebner, pers.~comm.).  Based on this and the $Q_{Ar}$ estimate 
given above, we thus derive an estimate of [Ar/O]=0.0058$\pm$0.0017
in Hale-Bopp's coma.  In what follows we assume that this ratio of production rates is
indicative of the coma's Ar/O abundance ratio.  The most recent cosmogonic (i.e., solar)
[Ar/O] abundance ratio gives [Ar/O]$_{\odot}$=0.00372 \citep{GS98}.  Figure 2 presents the
derived correspondence between the error-weighted average brightness $B$ of the two Ar
lines, and the quantity $Q_{Ar}$/$Q_O$, ratioed to its cosmogonic value.

Although we recognize the difficulty of connecting coma to nuclear abundances with
precision because one does not at present know exactly how gases are stored in the
nucleus\footnote {As a result, inferring absolute abundance ratios with high confidence
will likely remain problematic until cryogenic comet nucleus samples can be returned to
Earth.}, we interpret our results at their face value:  Hale-Bopp's coma appears to be
enriched in Ar relative to cosmogonic proportions by 1.8$\pm$0.5:1.\footnote {CO is almost
as volatile as argon.  It is intriguing that Hale-Bopp's coma is enriched in Ar (this
paper) and also has a notably high CO abundance \citep{Bea97}.}  However, we must note that
the error estimate given here derives purely from the EUVS MRES counting statistics.  Our
effective area calibration uncertainty (35\%) could reduce this ratio to $\approx$1.2, but
it is equally likely to increase the ratio to 2.4; so too, there is a dispersion in the
various Hale-Bopp $Q_{\rm H_2O}$ measurements at perihelion of 20\%, which could reduce
this ratio to $\approx$1.4, but it is equally likely to increase the ratio to 2.2, and a
quoted 16\% uncertainty in the solar Ar/O ratio \citep {GS98}.  Adding {\it all} of these
errors in quadrature, we find Hale-Bopp's coma appears to have an Ar/O ratio relative to
cosmogonic proportions that is 1.8$\pm$0.96:1.

Because noble gases by their nature do not participate in either the cometary ice chemistry
or coma chemistry, noble gases are parent species in the coma, and their abundance in the
neutral coma should reflect their bulk abundance within the cometary nucleus.  We assume this
is the case in what follows.\footnote {Though near-surface thermal segregation effects, which
are difficult to accurately predict at present, could in principle affect the accuracy of
this correspondence.}

The lack of depletion of Hale-Bopp's Ar relative to cosmogonic proportions is
interesting in several respects.  The first of these concerns the thermal history of
Hale-Bopp itself.  The fact that Ar is not depleted in Hale-Bopp indicates that the deep
interior of Hale-Bopp's nucleus cannot have ever reached equilibrium temperatures of
perhaps 35--40 K, else the Ar would have been lost to sublimation \citep{BO98}.  Neon,
which is even more volatile than argon, sublimates vigorously at temperatures of 16--20 K.
The EUVE satellite returned spectra of Hale-Bopp in 1996 showing Ne is $>$25 times depleted
relative to cosmogonic proportions \citep{Kea97}; this is strong evidence that Hale-Bopp's
deep interior {\it has} been warmed above 20 K.

Together, the EUVE satellite and EUVS rocket experiment results trap Hale-Bopp's bulk
internal temperature as having exceeded the Ne sublimation loss range of 16--20 K, but not
the Ar sublimation range of approximately 35--40 K.  These temperature constraints are
consistent with others Hale-Bopp retrieved from H$_2$O ortho-para \citep{Crea97} and D/H
ratios \citep{Bea99}.  We interpret our result as indicating that either the solar nebula
was far colder and far richer in Ar than models typically predict \citep{Lea00}, or that
Hale-Bopp was formed in the cold, Kuiper Belt region (i.e., well beyond the 20--30 AU
Uranus-Neptune zone) and was then subsequently ejected to the Oort Cloud (its more recent
dynamical home) without ever spending much time in the (warmer) Jupiter-Saturn zone, or
both. Although either implication is contrary to ``conventional wisdom,'' we
note that the transport of comets from the Kuiper Belt region (usually after
Neptune-induced inward evolution) to the Oort Cloud has been detected in recent
dynamical simulations of Oort Cloud formation.\citep{Dea00}.

This first detection of a native cometary noble gas, argon, whets our appetite
for more measurements, in order to assess how typical Hale-Bopp is in its Ar
abundance.  We also look forward to detection of other noble gases in comets,
particularly Ne and Kr, and to the eventual determination of whether the ices
within individual comets contain varying proportions of Ar and other chemically
inert tracers as a guide to understanding the heterogeneity vs.~homogeneity of
material within cometary nuclei, and therefore better still understanding their
origins.



\acknowledgments

We thank NASA mission manager Anel Flores, NASA Sounding Rocket branch, NASA Headquarters
discipline scientist Henry Brinton, and the entire NASA/contractor team that made these
time-critical observations a success.  At SwRI, we thank our support staff, including Tom
Booker, John McDonald, Clarence McGinnis, David Orth, Bill Tomlinson, and Roy Welch.  At
the University of Colorado, we thank Gary Kushner and Scott McDonald for assistance with
our instrument calibration.  Leslie Young provided a careful and insightful reading of this
manuscript; David Grinspoon and Walter Huebner provided illuminating discussions.  An
anonymous referee made a series of helpful comments as well.  This work was supported by
NAG5-5006.





\clearpage



\figcaption[hb_ar_fig1.eps]{{\it Left-Hand Panel:}  The spatially-collapsed 950--1100 \AA\
brightness spectrum of Hale-Bopp derived from the EUVS medium resolution slit (MRES) dataset
centered very close to the center of brightness in Hale-Bopp's coma; the data shown were
taken exclusively above 200 km altitude.  Both the H I Ly $\beta$ and the Ar I features
cited in the text are identified; the nominal instrumental pseudo-continuum background level
we adopted is also shown, as a dashed line.  {\it Right-Hand Panel:}  Comparison of the
Hale-Bopp spectra obtained in the EUVS medium resolution (MRES, bold line) and high
resolution (HRES, thin line) slits datasets, with the HRES dataset having been rebinned to
$\approx$6 \AA/bin; the presented offset and scaling of the two spectra in the righthand
panel was set purely for plotting clarity.\label{fig1}}

\figcaption[hb_ar_fig2.eps]{Derived Ar perihelion production rate estimate, compared to the
cosmogonically-predicted Ar production rate one would expect based on Hale-Bopp's total
oxygen production rate; the error bar shown here is due to counting statistics alone.  See
text for calculation, additional error bar details, as well as the interpretation of this
result.\label{fig2}}


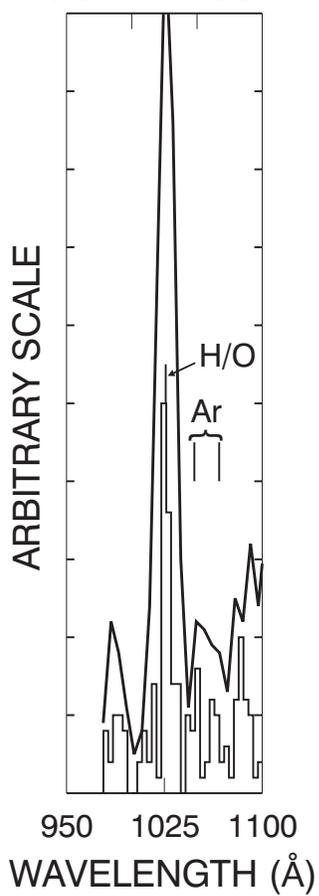
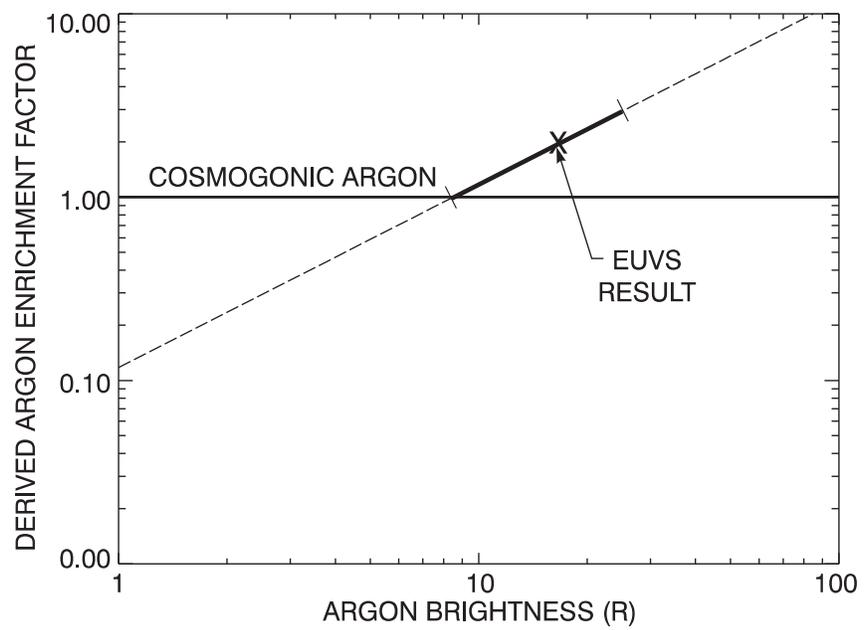
\end{document}